\documentclass[twoside]{article}

\usepackage{PRIMEarxiv}

\usepackage[utf8]{inputenc} 
\usepackage[T1]{fontenc}    
\usepackage{hyperref}       
\usepackage{url}            
\usepackage{booktabs}       
\usepackage{amsfonts}       
\usepackage{nicefrac}       
\usepackage{microtype}      
\usepackage{lipsum}
\usepackage{fancyhdr}       
\usepackage{graphicx}       
\graphicspath{{media/}}     

\usepackage[nointegrals]{wasysym}
\usepackage{algorithm2e}
\usepackage{caption}
\usepackage{subcaption}

\usepackage{amsthm}
\usepackage{amsmath}

\theoremstyle{definition}
\newtheorem{definition}{Definition}
\newtheorem{remark}{Remark}

\title{Skills Composition Framework for Reconfigurable Cyber-Physical Production Modules
}

\author{
  Aleksandr Sidorenko, Achim Wagner\\
  Deutsches Forschungszentrum für Künstliche Intelligenz GmbH (DFKI) \\
  Trippstadter Strasse 122, 67663 Kaiserslautern, Germany\\
  \texttt{\{Aleksandr Sidorenko\}aleksandr.sidorenko@dfki.de} \\
   \And
  Martin Ruskowski \\
  Deutsches Forschungszentrum für Künstliche Intelligenz GmbH (DFKI) \\
    Trippstadter Strasse 122, 67663 Kaiserslautern, Germany\\
Institute of machine tools and control systems (WSKL)\\
University of Kaiserslautern-Landau, 67663 Kaiserslautern, Germany\\
}

\begin{document}
\maketitle

\begin{abstract}

  While the benefits of reconfigurable manufacturing systems (RMS) are well-known, there are still challenges to their development, including, among others, a modular software architecture that enables rapid reconfiguration without much reprogramming effort. Skill-based engineering improves software modularity and increases the reconfiguration potential of RMS. Nevertheless, a skills' composition framework with a focus on frequent and rapid software changes is still missing. The Behavior trees (BTs) framework is a novel approach, which enables intuitive design of modular hierarchical control structures. BTs have been mostly explored from the AI and robotics perspectives, and little work has been done in investigating their potential for composing skills in the manufacturing domain. This paper proposes a framework for skills' composition and execution in skill-based reconfigurable cyber-physical production modules (RCPPMs). It is based on distributed BTs and provides good integration between low-level devices' specific code and AI-based task-oriented frameworks. We have implemented the provided models for the IEC 61499-based distributed automation controllers to show the instantiation of the proposed framework with the specific industrial technology and enable its evaluation by the automation community.

\end{abstract}


\section{Introduction}
\label{sec:intro}

The Industry 4.0 paradigm manifests a shift towards mass customization and beyond to personalized production. This requires new levels of flexibility, reconfigurability and intelligence from production lines and machines. While the benefits of the reconfigurable manufacturing systems (RMS) are-well known, they have not fully found their way to the production floor yet. One of the reasons is that the industrial software still cannot provide the required level of modularity and reconfigurability, while remaining reliable and safe. Another problem is a high level of complexity of such systems. That is especially challenging for small companies that cannot afford to hire versatile experts to maintain and (re-)configure such machines \cite{brunoeReconfigurableManufacturingSystems2017}.

Skill-based engineering, a popular approach in factory
automation, helps to increase the reconfiguration potential of the RMS by decoupling the required production processes from those that are executed on production resources, and encapsulating machine-specific functions behind standard interfaces \cite{dorofeevSkillBasedEngineeringIndustrial2023}. Although skill-oriented approaches have shown some promising results, we argue that a framework for skills composition with the special focus on machine's reconfigurability is still missing. Modelling and realization issues are among the main challenges \cite{froschauerCapabilitiesSkillsManufacturing2022b}. In \cite{sidorenko_using_2022} we have defined the requirements for such a framework, such as modularity, modifiable hierarchical structure, ease of composition, and diagnosability. We have proposed to use behavior trees (BTs) as a supervisory control framework for composition of skills in reconfigurable cyber-physical production modules (RCPPMs).

In this paper, we continue to investigate the potential of BTs for skills' composition in the manufacturing domain. We propose formal models of skills, their composition, and execution. The models are then implemented in the IEC 61499 based modelling environment to enable their evaluation by the automation community.

The paper is structured in the following way. \autoref{sec:stoa} gives an overview of current skill-based approaches both in manufacturing and robotics domains. In \autoref{sec: method} we describe our methodology and define the framework for skills composition and execution. \autoref{sec:results} discusses the resulted framework and shows an implementation example of the proposed models in the IEC 61499 4DIAC modelling environment. \autoref{sec:conclusion} concludes the paper with the future research plans.

\section{Related Works}
\label{sec:stoa}

\subsection{Skill-based engineering in manufacturing}

In their survey paper \cite{froschauerCapabilitiesSkillsManufacturing2022b} Froschauer et al. have listed the expected benefits of using skills in manufacturing, such as flexibility and planning efficiency in production, development efficiency, reuse, and interoperability. Among the main challenges, they named modelling, realization, planning issues, as well as lack of standardization and correct execution.

Koecher et al. have provided in \cite{kocherReferenceModelCommon2023} a reference model for common understanding of skills and capabilities in the manufacturing domain. The authors explicitly separate notions of capabilities and skills, as it allows to decouple description of a function from its implementation. While a capability is an implementation-independent specification of a function to achieve an effect in the physical or virtual world, a skill is “an executable implementation of an encapsulated (automation) function specified by a capability”. Skill's interface exposes its behavior through a harmonized state machine, as well as input and output parameters. The authors have also proposed an OPC UA based skill model, which utilizes modelling capabilities of the OPC UA technology. The model focuses mostly on the skill's invocation and leaves the implementation details to the developers.

In the last few years, several prototypes have shown the applicability and benefits of using skill-based design for modular, plug-and-produce capable production systems.
Models based on OPC UA \cite{volkmann_integration_2021, zimmermannSkillbasedEngineeringControl2019, profanter_generic_2021} are still mostly custom and rely on a lot of background programming. Skills implementation still needs a lot of manual work.

Dorofeev in \cite{dorofeevSkillBasedEngineeringIndustrial2023} proposes a holistic approach for skill-based engineering in the industrial automation domain. He stresses the relation of skills to services from service-oriented architectures (SOA) and focuses on the skills' orchestration. The execution code for an orchestrator is generated based on the so-called sagas pattern. He notes the advantage of code generation, especially when it comes to complex skill orchestration. A service contract is comparable to what we call a skill execution protocol. It is also not clear how the approach models the skill's execution context.

\subsection{Skill-based design in robotics.}

Skill-based design has long been used in the robotics domain, though the focus has shifted in the direction of reusability and modularity of code, as well as  separation of roles during development. Skills are reusable components, which encapsulate expert knowledge in some task, e.g, in trajectory planning. Skills normally reside on the lower levels of some layered architecture, e.g., three-layer architecture, as in \cite{schlegelNavigationExecutionMobile2004}, and are used by a sequencing layer to automatically build complex robot's behaviors. When a robot operates in changing environments, its behavior must adjust to the current situation. This requires the robot's skills to be more fine-granular than the machining skills. The focus lies more on the skill's composition framework than on the skill's standard interface.

An approach for task-level programming of flexible mobile manipulators in industrial environments has been proposed in \cite{boghDoesYourRobot2012}. The authors have defined the terminology of tasks, skills, and motion primitives, and shown how this can be used for separating responsibilities of the robot software developers. They have also introduced  the concepts of skill's pre- and postconditions.

Rovida et al. in \cite{koubaa_skirosskill-based_2017} has defined a robot skill as a fundamental software block, which changes the world state. An operator configures a robot by defining a scene and a goal, and the robot plans the required sequence of skills to achieve the goal. They also stress the necessity of skills' reuse and sharing.

\subsection{Skills for Reconfigurable CPPMS.}

Reconfigurable cyber-physical production modules (RCPPMs) have modular structure and can change their production capabilities and capacities to adapt to changing conditions. Such production modules can be considered as a further evolution of the RMS ideas with their application to module's components. One of the key requirements here is that such modules must be reconfigured rapidly without any programming efforts, following the principals of plug-and-produce.

Dorofeev notes in \cite{dorofeevSkillBasedEngineeringIndustrial2023} that skill-based design improves the software characteristics and the reconfigurability potential of RMS. Though, RCPPMs present special challenges to skills-oriented approaches. As submodules and components inside RCPPMs are closely integrated, this involves a lot of execution context and dependencies between skills. The information provided by the interface state machines is not enough for composing complex behaviors. That is why, a lot of custom programming is still needed for skills composition. Modularity (modular hardware and software system components) and integrability (integration of new components and their functions, which have not been seen in the system before) characteristics of RCPPMs imply the absence of a central controller that is aware of all its subordinate components. A central orchestrator approach will not work in that case. Instead, the control software should probably resemble the modular and hierarchical structure of RCPPMs. The challenge here is that such control structures must be rapidly reconfigurable without much programming.

In our skill-based design for RCPPMs, we follow both approaches, but focus more on the composition framework that defines the composition rules interfaces, which must be then implemented by all participants. The framework defined in this paper is fully compatible with the reference model presented in \cite{kocherReferenceModelCommon2023} and provides good synergies with future capability models.

\section{Methodology}
\label{sec: method}

The goal of the proposed models is to enable the rapid design of robust reactive behavior for the current configuration of a production module and the task given. If the task cannot be completed with given resources, then it should be able to adapt the module's configuration and behavior. The base assumption is that different components inside a production module provide a set of parametrisable controllers, which, given the right set of narrow conditions, guarantee to complete their tasks successfully. A system can intelligently switch between these controllers to produce the required behavior. The process of choosing the right controllers is decoupled from the low-level control policies, and thus can be rapidly adapted to the new module's configuration.

Authors in \cite{burridgeSequentialCompositionDynamically1999} proposed to compose controllers sequentially to produce complex control strategies, if one controller would “push” the system to the state where the next controller could catch up. They used a metaphor of control policy funnels that moved a larger set of system's initial conditions to a smaller set of final conditions, which characterized the goal. Oegren in \cite{Ogren2020} used behavior trees (BTs) as a supervisory control framework to generalize and improve the approach from \cite{burridgeSequentialCompositionDynamically1999}.

These ideas build a basis for our skills' composition framework. Each atomic skill can be viewed as a parametrisable controller, implemented as a special type of BT, and a composite skill is a composition of such BTs. The choice of BTs formalism for skills' composition is motivated by its modularity, hierarchical structure and ease of composition. Modularity of BTs enabled us to combine them with another systems' composition framework introduced by Reich in \cite{reichCompositionCooperationCoordination2021}. These two approaches complement each other and define our modular distributed hierarchical control framework for RCPPMs.

\subsection{Skill Model}
\label{subsec: Skill-Model}

We differentiate between descriptive and operational models of automation functions, as in \cite{nauBlendedPlanningActing2015}. Descriptive models help us to choose an appropriate action in the current situation to achieve our goal, while operational models describe how to perform the chosen action or actions and how these actions should be organized, given the current execution context. The reason to separate these models arises from the fact that they have normally different levels of abstraction and provide different information. For deciding to drill a hole in a part, we do not need to know how exactly to do it. Capabilities and skills from \cite{kocherReferenceModelCommon2023} are a good example of such separation. Though, as mentioned by authors in \cite{nauBlendedPlanningActing2015}, by using the same knowledge representation for both types of models it makes it easier to maintain consistency between planning and acting. We start with a formal definition of a skill, which is general enough to be used in both descriptive and operational models of CPPMs' functions.

Let the system state be \(x \in \chi \subset \mathbb{R}^n \). To represent a logical state of the system, we can use \textit{conditions C}, which are boolean-expressions over the state variables. A state \( x \) satisfies the condition \(C_j\) if \(C_j\) evaluates to \(True\) when all variables are assigned values according to the valuation \( x \).

\begin{definition}[Skill]
  \label{def: skill}

  A skill \(\mathbb{S}\) is a tuple:

  \begin{align}
    \mathbb{S} & = \{C_{pre}, C_{inv}, C_{post}, F_{skill} \},
  \end{align}

  where \(C_{pre}\) is a \textit{precondition}, which represents a logical state, where the skill's function can be started. \(C_{inv}\) is the skill's \textit{invariant}, a condition, which must be \(True\) all the time the skill is running. \(C_{post}\) is \textit{postcondition} and defines the expected logical state that results from the running of the skill. \( F_{skill} \) is the skill's function, which transforms the system's logical state from \(C_{pre}\) to \(C_{post}\).

\end{definition}

The elements in this formulation are intentionally very general. This leaves the room for various skill's types, e.g., a purely algorithmic skill, which calculates a trajectory, or a device's skill, which represents a controller. In the case of the controller, if \( x \) is the system's state, then \( F_{skill} \) is defined as \( x_{t+1} = f(x_t, u_t)\)  for discrete time, where \(u \in U \subset \mathbb{R}^k \)  is a control variable.

An interface \(I_{skill}\) adds the operational properties to the skill model. It provides all the information required for automatic use and composition of skills.

\begin{definition}[Skill Interface]
  \label{def: skill_interface}

  A skill's interface \(I_{skill}\) can be defined in the following way:

  \begin{align}
    I_{skill} & = \{A, Param, M \},
  \end{align}

  where \(A\) is a skill's interface I/O automaton, \(Param\) is a set of skill parameters, such that an optional mapping function \(M\) maps each skill's parameter to its function's inputs and outputs variables: \( p \in Param\mid Map(p) \rightarrow \ I \cup O \). \(I\) and \(O\) are two disjoint finite sets of typed input and output state variables. A mapping function \(M\) provides an abstracting functionality to the skill's parameters. For example, for a skill \textit{Pick} we might want to have an object as a parameter, rather than target coordinates and gripper's closing position.

\end{definition}

The intuition behind the skill definition, is that the skill represents its function. It is agnostic to the specifics of the underlying algorithm or control policy and provides the standard interface needed for function's parametrization and composition, as well as the mechanism of pre- and postconditions to support the generation of the composition strategy.

\subsection{Operational Skill Model}
\label{bt-skill}

The presented here refinement of the skill model from \autoref{def: skill} utilizes a state space formulation of BT, which was introduced in \cite{Colledanchise2017}.

\begin{definition}[Behavior Tree \cite{Colledanchise2017}]
  \label{def: bt}

  A BT \( \mathcal{T}_i \) is a tuple

  \begin{align}
    \mathcal{T}_i & = \{ f_i,r_i,\Delta t \}
  \end{align}

  where \(i \in \mathbb{N} \) is a tree index, \( f_i : X \rightarrow X  \) is a function, which runs when the BT is executing, and \( {r}_i : X \rightarrow \{ \mathcal{R}, \mathcal{S}, \mathcal{F}  \} \) is a function returning feedback information regarding applicability and the progress of execution, \( \Delta t \) is a time step.
  The feedback \({r}_i\) is interpreted as \textit{\textbf{success}}(\( \mathcal{S} \)), \textit{\textbf{failure}} (\( \mathcal{F} \)), or \textit{\textbf{running}} (\( \mathcal{R} \)). We can partition the state space into the following regions, corresponding to the feedback data:

  \begin{align}
    \small
    R_i = \{ x \mid r_i(x) = \mathcal{R} \} ,
    S_i = \{ x \mid r_i(x) = \mathcal{S} \} ,
    F_i = \{ x \mid r_i(x) = \mathcal{F} \}
  \end{align}

  The execution of BT is defined by the equation:

  \begin{align}
    \small
    x_{k+t}(t_{k+1}) = f_i(x_k(t_k)) ,
    t_{k+1} = t_k + \Delta t
  \end{align}

  where \( x_k \) is the system state at the moment \( t_k \)

\end{definition}

A BT can be created either as a hierarchical combination of other BTs, using the composition operators, such as \textit{\textbf{sequence}} or \textit{\textbf{fallback}}, or by directly specifying a BT function \( f_i(x) \) and a feedback function \( r_i(x) \). In the last case, a BT node is called an \textit{\textbf{action node}}. When the feedback function returns \( r_i(x) = \mathcal{R} \), then the system dynamics is defined by the function \( f_i(x) \). Another major type of BT node is called \textit{\textbf{condition}}. A condition is also a BT \( \mathcal{T}_i \), such that its running region is empty \( (\mathcal{R}_i = \emptyset) \). This means that the condition node does not control the system, but checks logical conditions and influences the execution of the other actions.

BTs can be further composed in more complex BTs by so-called composition operators, which are also called the \textbf{\textit{control flow nodes}}. Two major composition operators are a \(Sequence\) and a \(Fallback\). The composition rules, which are encoded in these operators, are defined by the feedback information coming from the composed BTs. Here, we provide a formal definition only for the \(Sequence\). All the other operators are defined analogously and can be found in \cite{Colledanchise2017}.

\begin{definition}[Sequence composition of BTs \cite{Colledanchise2017}]
  \label{def: sequence}

  Two or more BTs can be composed into a more complex BT using a \textit{Sequence} operator,

  \begin{align}
    \mathcal{T}_0 & = \textit{Sequence}(\mathcal{T}_1, \mathcal{T}_2),
  \end{align}

  where \( r_0, f_0 \) of the \( \mathcal{T}_0 \) are defined in the following way:

  \begin{align}
    \text{If  }  x_k \in S_1 :
    r_0(x_k) = r_2(x_k),
    f_0(x_k) = f_2(x_k) \\
    else :
    r_0(x_k) = r_1(x_k),
    f_0(x_k) = f_1(x_k)
  \end{align}

  \( \mathcal{T}_1 \) and \( \mathcal{T}_2 \) are called children of \( \mathcal{T}_0 \). The definition effectively says that the BT  \( \mathcal{T}_0 \) will give control to the \( \mathcal{T}_2 \) only if \( \mathcal{T}_1 \) returns success. The \( \mathcal{T}_0 \) succeeds only if all its children return success. That is why the operator is called \(Sequence\).

\end{definition}

Informally, the \(Fallback\) works in the opposite way. It gives control to the second child only if the first one has failed, and it fails only if all its children have failed. It keeps trying the alternatives, hence the name \(Fallback\).

\begin{table*}[!b]
  \caption{Major BT nodes types.}
  \label{tab:bt-types}
  \begin{tabular*}{\hsize}{@{\extracolsep{\fill}}lllll@{}}

    \hline
    Node type & Symbol & Succeeds & Fails & Running\\
    \hline

    \(Sequence\) &   \(\rightarrow\) &  if all children succeed &  if at least one child fails &  if one child returns Running\\
    \(Fallback\)  & \(?\) &  if one child succeeds &  if all children fail &  if one child returns Running\\
    \(Action\) &  \(\Box\) &  on completion &  if impossible to complete &  during execution\\
    \(Condition\) &  \fullmoon &  if true &  if false &  never\\
    \hline

  \end{tabular*}
\end{table*}

Table \ref{tab:bt-types} summarizes the behavior of the major BT nodes types, which we use in our model. It should be noticed that the BTs framework has been designed extensible through designing new control flow nodes. For instance, there also exist a parallel composition node and various so-called decorators, which can significantly change the behavior of the overall system.

\begin{definition}[Skill as BT]
  \label{def: bt-skill}

  An operational model of the skill \(\mathbb{S}\) from \autoref{def: skill} is a behavior tree, defined by the following formula:

  \begin{align}
    \label{formula: skill-bt}
    \small
    \textit{Skill}_{BT} = \textit{Fallback}(\mathcal{C}_{post}, \textit{Sequence*}(\mathcal{C}_{pre}, \textit{Sequence}(\mathcal{C}_{inv}, f_{skill}) ))
  \end{align}

\end{definition}

\begin{figure}[!h]
  \includegraphics[scale=0.7]{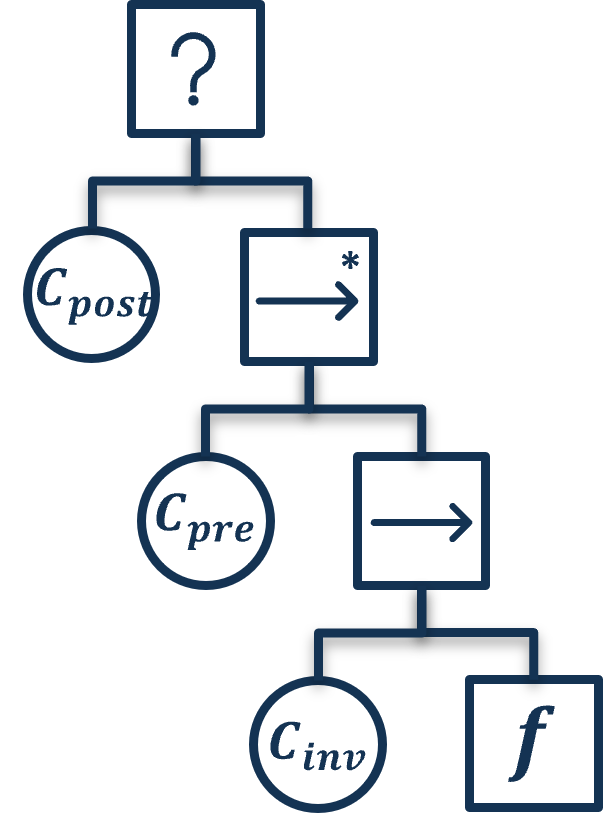}
  \centering
  \caption{Skill operational model as BT.}
  \label{fig:skill-bt}
\end{figure}

The graphical notation of the skill \( \textit{Skill}_{BT} \) as BT is shown in \autoref{fig:skill-bt}.

\begin{remark}

  The \(Sequence^*\) operator is called sequence with memory. It remembers the already succeeded children and does not consider their feedback statuses anymore. This is needed to prevent undesired switching back and forth between two control modes.

\end{remark}

The formula in \autoref{def: bt-skill} encodes the principals from \autoref{def: skill}. It can be interpreted as follows. The skill BT is called recursively. First, the \(Fallback\) checks the condition \( \mathcal{C}_{post} \). If it returns success \( x_k \in S_{postcond} \), then there is no need to continue, as the skill's goal has been already achieved. Otherwise, a second child of the \(Fallback\), the \(Sequence^*\) checks the condition \( \mathcal{C}_{pre} \). Normally, we want to check it just once, in the beginning. Returning success indicates that the system is in the state \( x_k \), from which the skill's function \( f_{skill}(x_k) \) can be started. The reactive sequence composition \( \textit{Sequence}(\mathcal{C}_{inv}, f_{skill})  \) ensures that the system does not violate some constraints during skill's function execution.

\subsection{Automatic Composition of Skills by Backchaining}

The proposed skill model enables us to use a technique called\textbf{\textit{ backchaining of BTs}} to automatically compose skills to complete a task. The backchaining of BTs was first proposed in \cite{colledanchiseBlendedReactivePlanning2019}. The approach is based on the two main ideas: first, creating an atomic BT for each action template in the precondition-postcondition-action form; second, iteratively replacing failed conditions with the atomic BTs, which work to make the failed conditions to succeed.

\begin{figure}[!h]
  \includegraphics[scale=0.7]{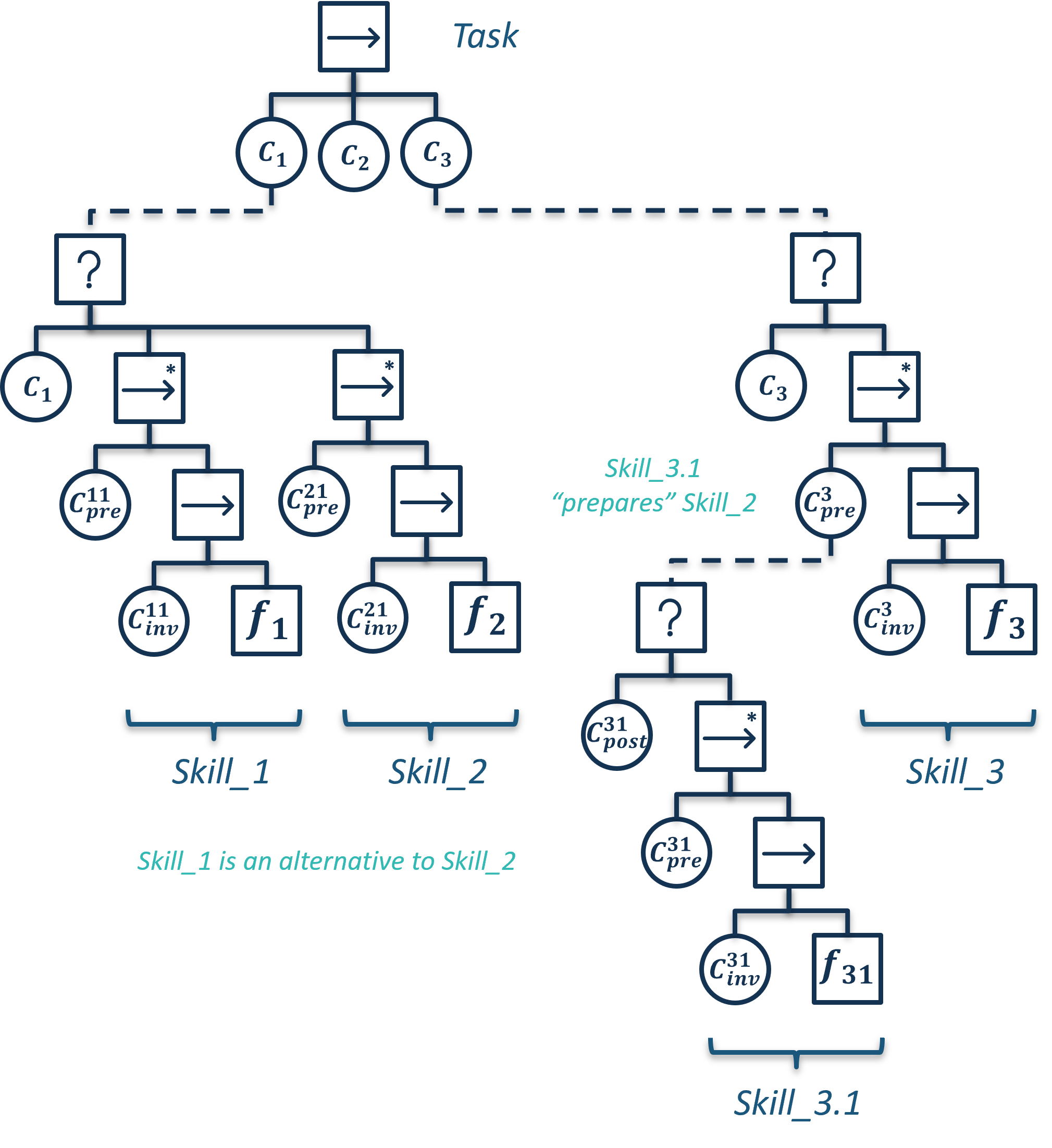}
  \centering
  \caption{Skills composition by the BTs backchaining.}
  \label{fig:backchaining}
\end{figure}

The BTs backchaining has some similarities to backward search algorithms from automated planning. Its working principal is shown in \autoref{fig:backchaining}. A goal or a set of goals is presented as a sequence composition of conditions  \( (C_1, C_2, C_3) \). The priorities go in the sequence from left to right, i.e., \( C_1 \) has the highest priority. If one encodes in this condition safety-critical goals, then they will always have the highest priority when choosing the course of action.
Now, assuming that a set of skills has been implemented and registered in the system, the task is to iteratively find those skills, which can achieve these conditions. For example, the \( Skill_1 \) can achieve the condition \( C_1 \) if its postcondition is the same, i.e., \( \mathcal{S}_{Skill_1} \subset \mathcal{S}_{C_1} \) (\( \mathcal{S}\) is success region). We will leave the question of how to find equal conditions out of this work's scope. The answer depends on the descriptive models of skills and conditions. There might be several skills that satisfy the required condition \( C_1 \) (\( Skill_1\) and \(Skill_2 \) in \autoref{fig:backchaining}). We can compose such skills using a fallback operator and replace the condition \( C_1 \) in the sequence by the resulting BT. This makes the resulting control strategy more robust to disturbances, as, if the \( Skill_1 \) fails, the \( Skill_2 \) tries to reach the condition \( C_1 \). It should be noted that in this type of the fallback operator the priorities of the children is fixed, i.e., the \( Skill_1\) has the higher priority as the \( Skill_2\) and will always be activated the first. It will be prudent to give higher priority to a skill that is more likely to succeed.

If a precondition of a skill fails, e.g., \( C_{pre}^3\) of the \( Skill_3 \) in \autoref{fig:backchaining}, then we can try to find another skill with the same postcondition \( C_{post}^{3.1}\) of the \( Skill_{3.1} \) and replace the failed condition with this skill.

\begin{figure}[!h]
  \includegraphics[scale=0.7]{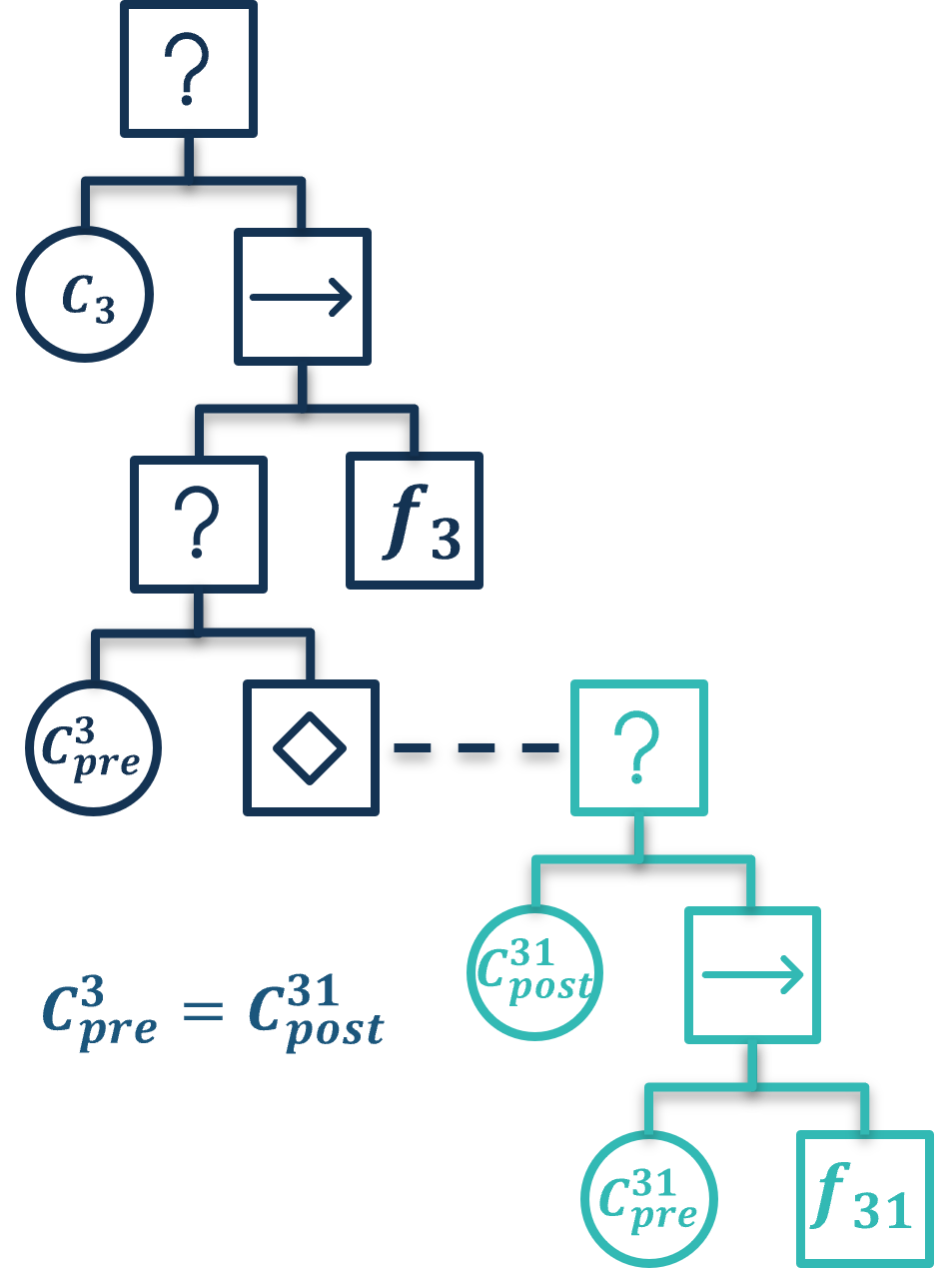}
  \centering
  \caption{A pattern for replacing a failed skill's precondition with a refinement BT.}
  \label{fig:bt-lookup-decorator}
\end{figure}

The backchaining method works in principle with some exceptions, which were shown and analyzed in \cite{Ogren2020}. Though, to apply it in the distributed control architecture of the RCPPMs, it requires some improvements. First, to find the required skills, one needs to compare pre- and postconditions. After the skill is found, its BT must replace the failing precondition. A possible solution pattern is schematically shown in \autoref{fig:bt-lookup-decorator}. We can use a fallback composition of the skill's precondition and the special decorator node (the node with the diamond shape in \autoref{fig:bt-lookup-decorator}), which will “synchronize” with the required subtree \( \mathcal{T}_{31} \) using the mechanism discussed in \cite{sidorenko_using_2022}. Decorators are the BT's custom control flow nodes with one parent and only one child, which alter the control flow in some custom way. They have been introduced to the BTs frameworks to systematically extend its functionality. Furthermore, skill's descriptive models will be needed to find the right skill and configure it. We see here a good interplay with the capability models.

\subsection{Executing Skills on Distributed Control Components}

As mentioned earlier, RCPPMs are modular and consist of various networked cyber-physical components running concurrently. We compose concurrently running processes using the approach, proposed by Johannes Reich in \cite{reichCompositionCooperationCoordination2021}.

\begin{figure}[!h]
  \includegraphics[width=0.7\textwidth]{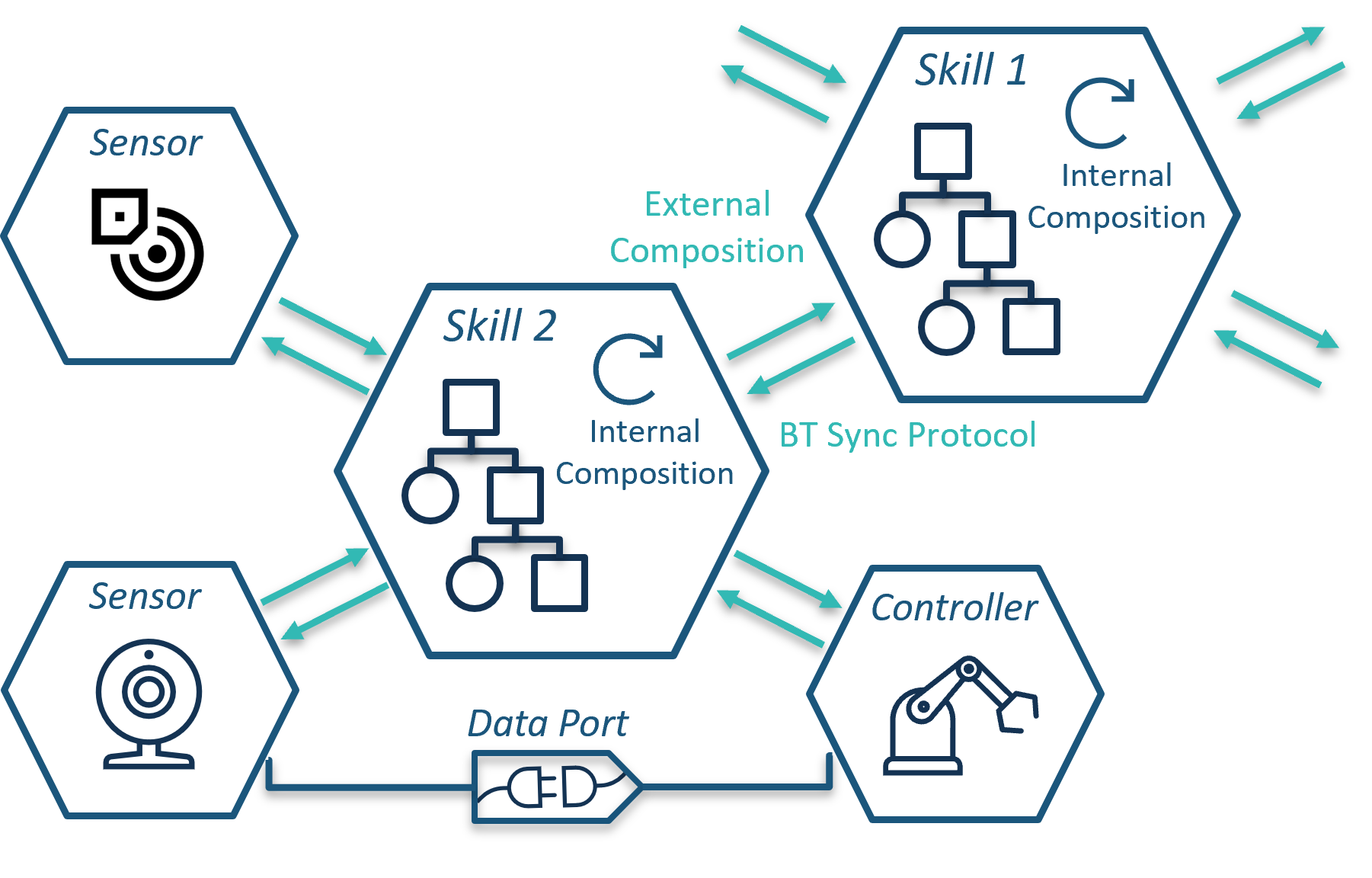}
  \centering
  \caption{Snowflake topology for the execution of BTs as distributed components.}
  \label{fig:bt-distributed-exec}
\end{figure}

Reich proposes to divide the composition of interactive systems into external and internal. The external composition takes place when so-called interaction partitions of each process communicate with each other through protocols. These partitions are called roles and are modeled by IO-automata. When two IO-automata are coupled by a protocol, the output of the first automation becomes the input of the second and vice versa. We can compose these IO-automata into a protocol automaton and eliminate all potential deadlocks, i.e., non-final states without continuation, and livelocks, i.e., periodically reached states without continuation to reach an acceptance condition. The protocol represented by such an automaton is called consistent. We have used this approach to design a \textit{skill execution protocol} in \cite{sidorenko_opc_2021} and a \textit{BT synchronization (BTSync) protocol} in \cite{sidorenko_using_2022}.

By the internal composition, Reich means the coordination of different roles inside each component. He has shown in \cite{reichCompositionCooperationCoordination2021} that if two roles from two different consistent protocols are coordinated in such a way that the respective acceptance conditions remain fulfilled, then the composite structure again results in a consistent protocol. Composed is this way, computation processes form a so-called \textit{snowflake} topology without any cycles.

To complete our skills' composition framework, we combine the skill model from \autoref{def: bt-skill} with the model proposed by Reich in \cite{reichCompositionCooperationCoordination2021} in the following way. If BTs are running on the separate components or in the separate thread of control concurrently, they are composed using the external composition by protocols. As the BTs model enforces the use of one interface across all its nodes, we need only one protocol, which has been defined in \cite{sidorenko_using_2022}. For the BTs running locally on one computation resource, we can use the composition rules defined in \autoref{bt-skill}. The internal composition of the interface state machines is done by a BT, running inside a computational process. The analysis done in \cite{Colledanchise2017} provides the criteria for a BT to converge to its success region in a finite time. Such BT is called a \textbf{\textit{Finite Time Successful (FTS)}}. It can be shown that if an FTS BT \( \mathcal{T_{FTS}} \) is used for the internal composition of two or more BTSync protocols, then the composed control structure is globally also consistent. The modular distributed control architecture based on the resulting snowflake topology is shown in \autoref{fig:bt-distributed-exec}. This topology supports the tree structure of BTs and enables to run it without the loops. We alloy connections between the tree leaves only for the data transfer through the typed data ports, as it does not break the BTs flow of control.

\begin{figure*}[!h]
  \includegraphics[width=1.0\textwidth]{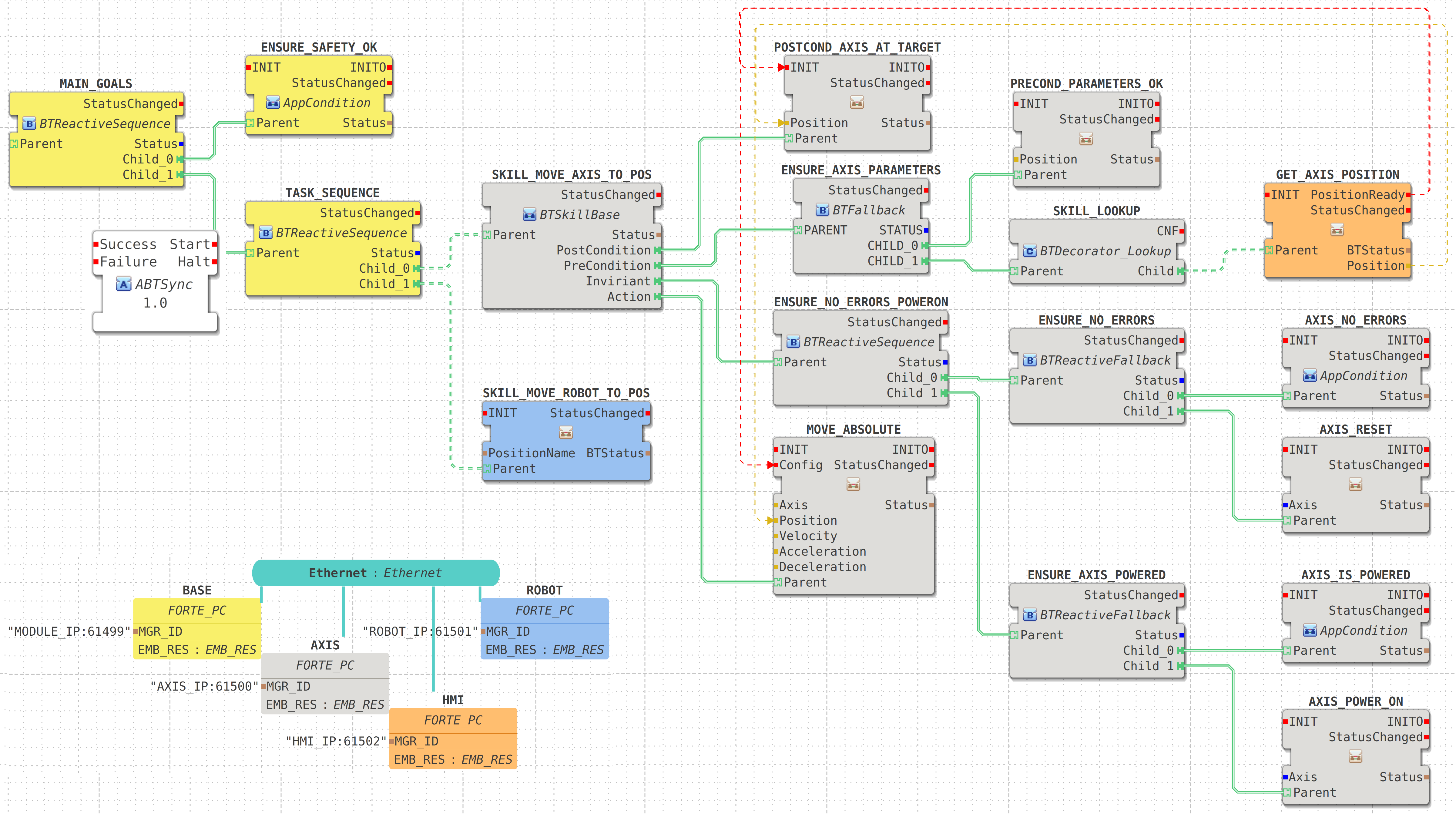}
  \centering
  \caption{Implementation of the framework in the 4DIAC.}
  \label{fig:4diac-moveto-skill}
\end{figure*}

\section{Implementation and Results}
\label{sec:results}

\subsection{Assessment of the Proposed Model}
\label{subsec:assessment}

The proposed model is aligned with the skill definition in \cite{kocherReferenceModelCommon2023}, but also provides several advantages. First, it packs the skill's execution context together with its skill in one operational unit (see \autoref{def: skill}). This localizes the skill's dependencies and improves the modularity and diagnosability of the design. Second, by using BTs as the skill's operational model, the skill's execution context is checked automatically. Skills, which describe their behavior through a simple state machine, do not provide enough information about their execution context. A system must make all the necessary checks before starting the skill itself. The BT from \autoref{def: bt-skill} makes all the required checks efficiently and robustly. This simplifies the composition rules. Third, the model enables a better alignment of the high-level descriptive models with the operational ones, as both can use the same knowledge representation. This allows to blend planning and acting and improve the integration between the low-level, device-specific functions and the high-level AI-based task-oriented frameworks.

The interface part, which is separated from the skill, makes the model more specific, as it defines how the components are composed together. The interfaces can be standardized, so the systems using them can rely on these standards. Alternatively, a composition framework can enforce the use of specific interfaces, which must be implemented by the components in order to participate in composition. In our work, we rely on this alternative approach. We started with the composition framework that provided us with the desired properties, defined by the requirements in \cite{sidorenko_using_2022}. The resulted framework has defined the skill interface (BTSync protocol) and the skills' composition rules (BTs composition operators). Thanks to the self-similarity property of BTs this interface and these composition rules are applied identically on all the levels of control hierarchy.

The approach has also shown the benefits of integrating BTs and state machines.
The snowflake topology and the concept of external and internal compositions enable a consistent distributed control architecture. The BTs constrain the external composition to using only one protocol, the BTSync protocol, defined by the BTs metamodel.
The BTs appear to be overlaid on the snowflake topology and create flexible hierarchical control structures that are distributed across different cyber-physical components and run concurrently. This resembles the physical structure of the RCPPMs. The resulting control hierarchies can be reconfigured in runtime due to the modularity of the overall architecture.


\subsection{Implementation}
\label{subsec:implementation}

This section provides a small example, which illustrates, how the models presented in \autoref{sec: method} can be implemented using specific technology, namely the IEC 61499 standard for distributed automation systems. In our recent paper \cite{sidorenkoUsingBehaviorTrees2024}, we have shown how BTs can be integrated with the state-of-the-art PLC technologies, IEC 61131 and IEC 61499-based systems. Here, we use the developed in \cite{sidorenkoUsingBehaviorTrees2024} event-based version of BTs to demonstrate the proposed skill-based design. The example was implemented in the open source 4DIAC \footnote{https://eclipse.dev/4diac/en\_ide.php} IDE.

\autoref{fig:4diac-moveto-skill} shows the BT of the application. The BASE component (yellow in \autoref{fig:4diac-moveto-skill}) builds the system's basis and provides all the system critical infrastructure. It is envisioned that this part will be deployed by the machine's producer and users will not get direct access to it. As this part of BT has the highest priority, the safety critical functions are located here. The AXIS and the ROBOT components are controlled separately by their own BTs. The task is a simple sequence of moving the axis and then the robot to their target positions.
The AXIS BT (gray in \autoref{fig:4diac-moveto-skill}) shows the implementation of the models presented in the paper. First, one can see the snowflake topology of the BT design.
Each node of the BT is implemented as a function block (FB) and is plugged into the tree through the Adaptor, which implements the BT synchronization protocol (the green lines connecting all the FBs). Control flows from left to right and from up to down, according to the tree structure of the BT.
The \textbf{skill\_move\_axis\_to\_pos} FB implements the skill BT formula from \autoref{formula: skill-bt}. The precondition \textbf{ensure\_axis\_parameters} shows the pattern from \autoref{fig:bt-lookup-decorator}. If the target position is not set, then the task of the \textbf{skill\_lookup} decorator will be to connect the skill, which provides the position. The skill \textbf{get\_axis\_position} runs on the separate hardware (orange HMI device in \autoref{fig:4diac-moveto-skill}) and asks an operator for the target position. After the operator confirms the position, the skill returns success and provides the required data to the \textbf{move\_absolute} FB, which implements the BT function. As mentioned, the procedure of finding and connecting the new skill to the tree has been left out of the paper's scope. The skill invariant condition \textbf{ensure\_no\_errors\_poweron} is refined by the small subtrees using the backchaining technique from \autoref{fig:backchaining}. They automatically reset the axis and power it on (if needed) when an error happens. The result BT is running on four separate control devises and can be easily reconfigured by introducing the new FBs and connecting them by the BTSync adaptor.

\section{Conclusion}
\label{sec:conclusion}

In this work, we have presented a framework for skills composition in modular reconfigurable control applications. First, we have provided a general formal skill model, which can be used as a skill metamodel for various descriptive and operational skill models. We have defined our skills' composition framework on top of the BTs formalism and composition of computational processes through protocols. The resulted control architecture is modular, hierarchical, self-similar, and allows rapid changes without much reprogramming. By using BTs as the basis, it allows good integration with AI technologies. It can provide a base operational layer for skills and enables further movement towards self-reconfigurable and adaptable skill-based CPPMs.

As we have focus in this work on the operational aspects of the model, to enable self-(re)configuration capabilities a good integration with descriptive models and AI-based technologies is needed. The BTs backchaining algorithm can serve as a basis, but needs to be improved. We see here a good synergy with the future capability models, as well as other descriptive frameworks. Furthermore, the proposed model needs a holistic approach for execution. Holonic architectures with their holarchies seem perfect candidates, and next we plan to investigate the integration possibilities. Finally, the OPC UA communication framework, especially the specifications from the OPC UA FX series, provides excellent opportunities for integration with the industrial automation landscape.

\section*{Acknowledgments}
This research has been supported by the European Union’s HORIZON Research and Innovation Program under the grant agreement No 101120218, the project \textbf{HumAIne} \footnote{https://cordis.europa.eu/project/id/101120218}, and conducted for the \textbf{Production Level 4} test bed environment at \textbf{SmartFactoryKL}\footnote{https://www.smartfactory.de/2022-production-level-4-oekosystem/}.

\bibliographystyle{unsrt}
\bibliography{references}

\end{document}